# What we talk about when we talk about monads

Tomas Petricek[a]

a   The Alan Turing Institute, London, UK

**Abstract**   Computer science provides an in-depth understanding of technical aspects of programming concepts, but if we want to understand how programming concepts evolve, how programmers think and talk about them and how they are used in practice, we need to consider a broader perspective that includes historical, philosophical and cognitive aspects. In this paper, we develop such broader understanding of monads, a programming concept that has an infamous formal definition, syntactic support in several programming languages and a reputation for being elegant and powerful, but also intimidating and difficult to grasp.

This paper is not a monad tutorial. It will not tell you what a monad is. Instead, it helps you understand how computer scientists and programmers talk about monads and why they do so. To answer these questions, we review the history of monads in the context of programming and study the development through the perspectives of philosophy of science, philosophy of mathematics and cognitive sciences.

More generally, we present a framework for understanding programming concepts that considers them at three levels: formal, metaphorical and implementation. We base such observations on established results about the scientific method and mathematical entities – cognitive sciences suggest that the metaphors used when thinking about monads are more important than widely accepted, while philosophy of science explains how the research paradigm from which monads originate influences and restricts their use.

Finally, we provide evidence for why a broader philosophical, sociological look at programming concepts should be of interest for programmers. It lets us understand programming concepts better and, fundamentally, choose more appropriate abstractions as illustrated in a number of case studies that conclude the paper.

# The Art, Science, and Engineering of Programming

**What we talk about when we talk about monads**

# 1 Introduction

*"A monad is just a monoid in the category of endofunctors. What is the problem?"* This quote has been intended as a joke [25], but it, surprisingly well, captures the paradoxical nature of how we talk about monads. Monads have a precise and succinct category theoretical definition that formally defines exactly what a monad is. Yet, this definition does not help you understand monads. It does not explain how and why they are useful in programming and, in some cases, excludes use cases that the community considers as monads. The quote also alludes to the belief that formal mathematical treatment of a programming concept leads to an ideal kind of knowledge about programming. Once you know that a monad is a monoid in the category of endofunctors, you know all there is to know.

You might want this paper to answer the question "What is a monad?", but that would make it just another monad tutorial. Instead, I will answer the question you should have asked instead, "What do authors really say when they talk about monads?" The answer situates monads, and writing about monads, within the disciplines of cognitive science, philosophy of science and mathematics. This paper reviews some explanations of monads, but it *does not* explain monads. Instead, it helps you gain a deeper understanding using a number of important, and often neglected, perspectives.

- We look at how monads are taught (Section 2) from the perspective of cognitive sciences. This teaches us that metaphors play an important role in understanding monads and are not a kludge employed by bad monad tutorial writers.
- We review how monads are used in academic literature (Section 3) and use the idea of research paradigms from philosophy of science to explain why some questions are asked and some are ignored.
- Looking at how monads are used by programmers in practice (Section 4), we trace how the notion of monad evolves. Although the formal definition stays the same, the uses of monads and the intuition behind them shifts and the meaning of a monad changes in a broader socio-technical sense.
- We briefly consider how monads are talked about in the community of programmers (Section 5). This provides an additional evidence that the narrow technical perspective does not explain why and how monads are used.
- We conclude by looking at three concrete cases where monads were used in a way that later publications dismissed as incorrect (Section 6). We argue that those could have been prevented if the metaphorical, philosophical and social understanding of monads, presented in this paper, was taken into account.

More generally, this paper provides foundations for a broader understanding of programming concepts that integrates formal and technical views with reflections obtained through cognitive sciences and philosophy. We argue that this is important not only to get a more complete understanding, but also for avoiding concrete problems when using programming concepts in practice.

## 2 Teaching and understanding monads

We start by briefly reviewing how monads are presented in academic literature and programming tutorials. The focus is on exemplifying how authors typically talk about monads, rather than on introducing monads in this paper. The side notes that authors make about monads reveal how monads are understood and the subtle transitions between mathematics and programming hint at the research paradigm in which much of work involving monads is grounded.

A typical introduction to monads links the concept to its formal origin. We follow this narrative in the next section and start with category theory. It is worth noting that this is what historians call an "internal history" – a history told within the discipline of computer science. A historian would aim to trace exactly how much influence category theory had on the development of the concept of a monad in programming. In contrast, a computer scientist might emphasize the historical link with category theory as an additional argument for a broader importance of the concept.

### 2.1 From category theory to programming

Monads were introduced by Godement [17] in algebraic topology in 1958. They became referred to as *standard construction* and *triples* before Saunders MacLane [34] popularized the name *monad* and also introduced the phrase *"a monad is a monoid in the category of endofunctors"*. The following definition of a monad is taken from a paper by Moggi [40] that first used monads in the context of programming languages:

**Definition 1.** *A* monad *over a category $\mathscr{C}$ is a triple $(T, \eta, \mu)$ where $T : \mathscr{C} \to \mathscr{C}$ is a functor, $\eta : Id_{\mathscr{C}} \to T$ and $\mu : T^2 \to T$ are natural transformations such that:*

$$\mu_A \circ T\mu_A = \mu_A \circ \mu_{TA}$$
$$\mu_A \circ \eta_{TA} = id_{TA} = \mu_A \circ T\eta_A$$

To a reader not familiar with category theory, the definition is very opaque. Understanding the details is not necessary for this paper, but it helps to think of the functor $T$ as a polymorphic type (such as a list), natural transformations $\eta$ and $\mu$ as functions and $T^2$ as a nested polymorphic type (list of lists of things).

It is worth discussing the original context in which the definition appeared. Monads were used to embed an object into an object with a richer structure [1] and also to express many different constructions in terms of the same structure [36]. Both of these are among the motivations for using monads in programming.

Moggi [40] used monads in semantics of programming languages to capture many different *notions of computation* that go beyond total functions, such as non-determinism, side-effects and exceptions. His paper provides categorical definitions for several monads later used in programming, however, it does not introduce them as programming tools, but rather as proving tools: *"This paper is about logics for reasoning about programs, in particular for proving equivalence of programs."*

Wadler [57] is the first to use monads as programming tools. Interestingly, similar structure was independently proposed by Spivey [50] without a reference to monads.

**What we talk about when we talk about monads**

Wadler translates categorical definition to a functional language. A functor becomes a type constructor (*operator on types*) with a *map* function and natural transformations become functions.

**Definition 2.** *A* monad *is an operator M on types, together with a triple of functions:*

$$\begin{aligned}
map &\ ::\ (x \to y) \to (Mx \to My) \\
unit &\ ::\ x \to Mx \\
join &\ ::\ M\ (Mx) \to Mx
\end{aligned}$$

*satisfying the following laws:*

(1) $join \circ join = join \circ map\ join$
(2) $join \circ unit = id = join \circ map\ unit$
(3) $map\ id = id$
(4) $map\ (g \circ f) = map\ g \circ map\ f$
(5) $map\ f \circ join = join \circ map\ (map\ f)$
(6) $map\ f \circ unit = unit \circ f$

The natural transformations $\mu$ and $\eta$ are now the *join* and *unit* functions, respectively. The first two monad laws (1-2) are direct translation of the laws in Definition 1. The remaining laws (3-6) are properties of natural transformations and so they are made explicit here.

The paper translates a number of monad definitions by Moggi into the language of functional programming, links monads with Haskell comprehension syntax and generalizes comprehensions from lists to any monad. We return to the topic of syntax in Section 4. The mapping is very direct, but it ignores the fact that programs are not guaranteed to terminate and so, strictly speaking, the programming language definition of a monad is not a correct category theoretical monad.

## 2.2 Bridging multiple levels of meaning

The early development that established monads as a programming concept hints at one important property of programming concepts that is central to this paper. The meaning of a programming concept such as monad comprises aspects at three different levels. The first is a mathematical aspect that can be used for formal analysis. The second is an implementation of the concept that is used for some purpose in concrete programs or libraries. Finally, the third is a metaphorical or intuitive understanding what the concept is in terms of general examples or analogies. In case of monads this is, for example, the idea that *"[monads] take some uninteresting object and turn it into something more structured"* [1].

Importing monads from mathematics into program code is also an interesting step philosophically. We use mathematics as a model of programs. To better structure our proofs in the model, we use a certain device provided by mathematicians. This device is then imported back into the programming world – going in the opposite direction than was originally intended when creating formal model of programs.

A *mathematical monad* and a *programming monad* are related, but they are not the same. In fact, philosophers would say that they are of a different kind [10], the first being *a priori* and the second being *a posteriori* kind of knowledge. Some aspects of what a monad is change as they bridge the gap. Moggi introduced monads for *reasoning* about effectful programs, but Wadler uses them to *implement* effectful programs in a purely functional programming language. Even a very direct translation subtly changes the purpose of a monad. Another change is that Moggi used *strong monads* with an additional operation that is not needed in programming (because of how programming languages handle variables) and so this aspect is omitted when we use monads in programming.

## 2.3 How programmers talk about monads

In programming, an alternative, but equivalent, definition of monads became popular shortly after the original appeared. The following is adapted from Wadler [58].

**Definition 3.** *A* monad *is a triple* $(M, unit, \ggg)$ *consisting of a type constructor M and two operations of the following types:*

$$
\begin{aligned}
\ggg &:: M x \to (x \to M y) \to M y \\
unit &:: x \to M x
\end{aligned}
$$

*These operations must satisfy the following laws:*

(1) $unit\ a \ggg f = f\ a$
(2) $m \ggg unit = m$
(3) $(m \ggg f) \ggg g = m \ggg (\lambda x.f x \ggg g)$

The $\ggg$ operator is known as *bind* and was initially written with the first two arguments switched and referred to as $\star$. The $\ggg$ version is easier to program with in Haskell, while $\star$ is easier to explain, so it is worth showing both:

$$
\begin{aligned}
\star &:: (x \to M y) \to (M x \to M y) \\
\ggg &:: M x \to (x \to M y) \to M y
\end{aligned}
$$

The $\star$ operation takes a function of type $x \to M\ y$ with monadic output and lifts it into a function $M\ x \to M\ y$ where both the input and the output are monadic; *unit* turns a value $x$ into a monadic value $M\ x$. Providing a more useful intuition about the meaning of the operations requires using a metaphor, which is what a typical monad tutorial does. We discuss such metaphors in the rest of this section.

Note that $M$ is now called a *type constructor* rather than a functor or operator on types. This is because the definition is now embedded in a language that supports polymorphic (generic) types that are used in the implementation. In a way, this makes the definition more narrow – monads could be encoded in many other ways and could be used in languages without polymorphic types, but the established definition heavily relies on this language feature and static types in general.

The programming literature on monads interprets and explains the formal definition either as purely formal or using one of two metaphors. Some authors treat such metaphorical explanations as kludges that are only needed because the concept is difficult to explain otherwise [61]. We argue that this is not the case and the metaphors are an important part of what a monad is. Before explaining why, the rest of this section reviews the three explanations of monads.

**Monad as a formalist entity**   The implementations of *bind* and *unit* for a concrete monad can perform a range of different things. Some authors, such as Wadler [58], try to avoid interpreting what the *bind* and *unit* operations of a monad represent and only describe concrete examples.

This view is akin to *formalism* as advocated by Curry [7]. According to formalists, mathematical statements have no inherent meaning and are just syntactic forms. Proving a property involves no understanding, but only an application of string transformation rules. However, avoiding interpretation and metaphors might be merely a way of writing – a style preferred by the mathematically oriented part of the programming community. As noted by Bayma in an interview quoted by MacKenzie [35]:

> *The way mathematicians write and talk are two different things. We write very proper, formal, very abstract. We think informally, intuitively. None of that is in the publication. When we get together we ask, "What does that mean?" (…).*

This suggests that, treating monad as a formalist entity is only done in writing, but when thinking and talking about monads, even a formalist uses metaphors. In writing, those are left out either for lack of space, or due to the belief that formal mathematics is a more pure, ideal form of knowledge (a topic we revisit in Section 3.2).

**Monad as a container**   The first metaphor is to see the monad $Ma$ as a container containing values of type $a$ [13]. This is a concrete version of the intuitive understanding in mathematics (Section 2.1) where a monad *embeds an object into an object with a richer structure*. Using this metaphor, we can explain a number of standard monads:
- List monad – $Ma$ is a list of $a$ values, or a container storing values in the list.
- Maybe monad – $Ma$ either contains a value of type $a$ or is an empty box.
- Reader monad – $Ma$ contains a value of type $a$ together with some other things.

Tutorials using this metaphor often use illustrations to intuitively explain what the abstract operations of a monad do [2]. When drawing a box to represent $Ma$, we follow some tutorial authors and draw multiple $a$ values. This gives a useful intuition about the *map* and *join* operations, but it can mislead us to think that the container can always contain multiple values – even though the metaphor can also be used for containers that contain exactly one or zero or one values.

$$
\begin{aligned}
\textit{unit} &: a \longrightarrow \boxed{a} \\
\textit{map} &: a \longrightarrow b \implies \boxed{a\ a} \longrightarrow \boxed{b\ b} \\
\textit{join} &: \boxed{\boxed{a}\boxed{a}} \longrightarrow \boxed{a\ a} \\
\textit{bind} &: a \longrightarrow \boxed{b\ b} \implies \boxed{a\ a} \longrightarrow \boxed{b\ b\ b\ b}
\end{aligned}
$$

According to the metaphorical explanation, the *unit* operation takes a value and wraps it in a box. The *map* operation takes a function that turns $a$ into $b$ and applies it to all things in a box. The *join* operation takes a box of boxes and unwraps it into a single box containing everything that was previously in the inner boxes. The *bind* operation is better explained as a combination of *map* and *join*. It takes a function that produces a box, applies it to all values in a box and then unwraps the nested boxes.

**Monad as a computation** The second metaphor is to see a monad as a computation [12]. Explanations using this metaphor typically use functions $a \to M\,b$, although the monadic value $M\,a$ can itself be seen as a computation. Monadic functions represent computations that are, in some way, non-standard, i.e. not naturally supported by the programming language. This can mean side effects in a pure functional language or asynchronous execution in an impure functional language. An ordinary function type cannot encode such computations and they are, instead, encoded using a data type that has the monadic structure.

This way of talking about monads focuses on composition of monadic computations through the *bind* operation, rather than on explaining what the monad itself represents, so it is complementary to the monad-as-container metaphor. Ordinary functions such as $f : a \to b$ and $g : b \to c$ can be composed using function composition $g \circ f : a \to c$. We use bold font face for the type $M\,a$ and write non-standard computations as as $a \to \mathbf{b}$. Note that, the input structure is now incompatible with the output structure, so functions like these cannot be directly composed using $\circ$.

The *unit* function is a computation $a \to \mathbf{a}$ that does nothing – it takes a value and returns it in a non-standard way without actually doing anything non-standard. The *bind* operation transforms a computation so that it can be composed into a bigger non-standard computation. Consider three functions:

$$f : a \to \mathbf{b} \qquad g : b \to \mathbf{c} \qquad h : c \to \mathbf{d}$$

The *bind* operation turns a function with non-standard output into a function that also has a non-standard input and propagates its non-standard aspect. We use the $\star$ version of the bind operation and write $f^\star$ for the result of applying bind to $f$.

$$f : a \to \mathbf{b} \qquad g^\star : \mathbf{b} \to \mathbf{c} \qquad h^\star : \mathbf{c} \to \mathbf{d}$$

Now the functions have compatible inputs and outputs and so we can compose them using $h^\star \circ g^\star \circ f$ to obtain one non-standard computation of type $a \to \mathbf{d}$.

One of the tutorials that follows this idea [60] makes the metaphor even more concrete and explains computations $a \to b$ in terms of railway tracks from the point $a$ to the point $b$. A non-standard monadic computation is a railway switch, so the three functions can be visually represented as follows. Rather than using bold face, the non-standard computation now has a straight track and a side track:

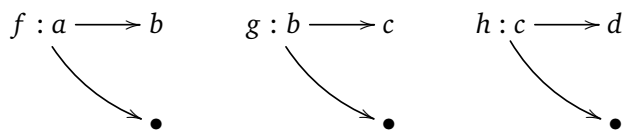

The fact that the railway tracks do not link corresponds to the fact that the functions cannot be composed. The bind operation represents an adapter that transforms a track with a switch into a two-track so that the railway can be composed:

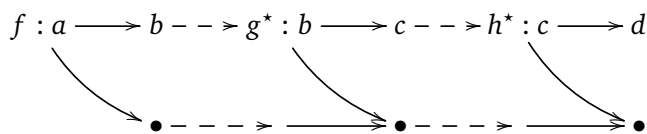

The railway switch is a good metaphor for some of the standard monads:

**What we talk about when we talk about monads**

- Maybe monad (again) – The direct track represents a computation that produces a value, while the side track represents a computation that produced no value.
- Exception monad – Similar, but the side track now represents a computation that failed and produced an exception instead.

Several other monads such as the continuation monad and the resumption monad [20] can also be seen as creating non-standard computation (one that returns asynchronously and one that can be stopped), but are not easily visualized as a railway track. However, for the two common cases where the railway track metaphor fits, it provides an effective way of explaining the concept.

## 2.4 Metaphors as an inseparable component

We typically see metaphors as a literary device. In the context of monad tutorials, they are seen as a kludge that aims to make understanding easier. Some argue [61] that this makes it hard to understand what monads really are and metaphors should be avoided in monad tutorials. However, research in cognitive science suggests the exact opposite. To quote Lakoff and Núñez [31]:

> *One of the principal results in cognitive science is that abstract concepts are typically understood, via metaphor, in terms of more concrete concepts.*

In other words, mathematical ideas are often grounded in everyday experience and metaphors link the everyday experience with more abstract concepts. Moreover, most thought is unconscious and inaccessible to introspection, so we are not directly aware that we are thinking in terms of metaphors.

One way to discover the metaphors behind our understanding is to look at the language we use to talk about abstract concepts [30]. This is also the case for monads. For example, the following are common phrases used when discussing monads:

- "In a monad" or "wrapped in a monad" is used to talk about the parameter $a$ of a generic type $M\,a$. This is the language we use for the container metaphor.
- The *unit* operation is also called *return*, suggesting the computation metaphor – the operation creates a computation that does nothing and returns the value.
- The name of the *bind* operation is also significant as it suggests let binding which is used for sequencing computations. The *bind* operation serves similar purpose when we think of monads as non-standard computations.

In summary, the two common metaphors used when explaining monads are more than just two concrete examples that are useful for understanding what a monad is. They are an inherent part of what a monad is to a programmer – both consciously and unconsciously. They provide a concrete concept, with attached everyday experience, that lets us think about monads via cognitive metaphors.

Even a pure formalist treatment of monads – as syntactic forms with no inherent meaning that are manipulated through string operations – can be understood in terms of metaphors. The hidden metaphor is *movement*. When we say, "in the next step [of a proof]", we are expressing ourselves as if we were talking about moving along a path.

## 3 Using monads in programming research

Since their introduction into programming language research, monads became a popular tool in academic work on programming language theory and in programming practice. We return to programming practice in the next section (Section 4) and briefly discuss how are monads treated in academic literature.

### 3.1 Reasoning about programs using monads

We do not aim to present a comprehensive literature review. Instead, we look at one particular aspect of monads, namely the monad laws. Those were mentioned by the definitions in Section 2, but where do they come from? Are they the right ones and what does it mean for the laws to be right? And what does the treatment of monad laws reveal about how we think about programming and computer science?

**Origins of monad laws** Algebraic laws about programming language constructs, such as those introduced by Hoare [21], are typically motivated by the aim to prove certain properties. If we have two programs that should intuitively be the same, we define a law that formalizes this requirement. For example, we might intuitively expect that if $b$ then $p$ else $p$ has the same meaning as $p$ and introduce the equality as a law [22].

Monad laws are not motivated by intuition about programs, but are, instead, imported from the category theoretical definition. The laws do not appear obvious when expressed in terms of *bind* and *unit*. They are better understood in terms of composition of monadic computations (following our second metaphor). Given monadic computations $f : a \to M\ b$ and $g : b \to M\ c$, we define a monadic composition $g \circ f : a \to M\ c$ as $\lambda x.f\ x \ggg g$. The laws then guarantee the following properties:

$$f \circ (g \circ h) = (f \circ g) \circ h$$
$$unit \circ f = f = f \circ unit$$

Those are the laws of a monoid and they guarantee that "*a monad is a monoid in the category of endofunctors*". However, this is not how monad laws are presented in the programming context and so they may appear somewhat ad-hoc.

Even though they do not originate from intuition about programs, monad laws are unanimously accepted by the programming community. In contrast, various extensions of monads [41, 45] usually motivate the laws by programming intuition, but there is often little agreement on the laws. The best example is the `MonadPlus` extension that adds an operation of type $M\ a \to M\ a \to M\ a$. There are at least two suggested sets of laws based on two different intuitions about the operation [14, 54].

**Reasoning with monad laws** Using monads for reasoning about programs was the original motivation by Moggi [40], but the immediate follow-up work used monads in a different way. Monad laws have later been used, as originally intended, for reasoning about programs with effects. For example, Gibbons and Hinze [15] use monad laws, together with laws about concrete monads, to reason about programs involving non-determinism, state and exceptions.

**What we talk about when we talk about monads**

However, using monad laws to prove program properties is less ubiquitous than the early work might suggest. More often, papers that refer to monads define a library for a concrete problem such as parsing, concurrency or parallelism [4, 24, 37] that fits the monad structure and satisfies the laws. Such papers often include a reference to monads in their titles, but their novelty is in solving a concrete programming problem.

It appears that monad laws matter less than the early work envisioned. Monad tutorials say that the purpose of the laws is to enable reasoning, but they rarely demonstrate such reasoning with a concrete example. There are also no programming tools offering, for example, refactoring of programs based on monad laws. There are refactoring tools that use monads [33] and that introduce monads [9], but not tools that are directly enabled by the monad laws. The use of monads for program reasoning remains a theme for an academic community that monads have since outgrown.

### 3.2 Monads and research paradigms

An established philosophy of science idea is that scientific research is grounded in a *research paradigm* [27] which provides a framework within which normal science is done. This includes methods of inquiry and the kind of questions that can be asked. At a smaller scale, *research programme* [29] is shared by a part of the community.

How and what we study using monads reveals the implicit assumptions of our current programming language research paradigm. In computer science, monads are rooted in a research programme that aims to *"utilize the resources of logic to increase the confidence (...) in the correctness of a program"*. Following Priestley [48], we refer to this as the *Algol research programme*.

Using monads is a way of subscribing to a certain research programme. The aforementioned papers that solve parsing, concurrency or parallelism problems [4, 24, 37] could treat monads merely as an unimportant implementation artifact, but monads are an accepted tool of the research programme and so a solution based on monads is often preferred. In Section 6, we will see that this may be problematic.

One of the core assumptions of the Algol research programme is that mathematics provides a fundamental source of knowledge about programming. This is what makes category theory a valid source of laws, even though the container and computation metaphors might suggest different laws.[1] Monad tutorials also follow the Algol research programme and introduce monad laws even if they do not use them later.

Equally interesting are questions that are rarely asked. In practice, many programmers struggle to understand monads because of their abstract nature. Some authors discuss how we could explain monads better, but the question whether it is desirable to use a monad (as opposed to a concrete notion) is often neglected. The assumption that abstraction is desirable is another part of the Algol research programme. Similarly, there are only few papers discussing syntax for using monads [16, 46] and those focus on formal properties of the notation, rather than on its cognitive aspects [3].

---

[1] For example, things stored in a container are not ordered, so we could argue that monadic bind should be commutative (i.e. $a \ggg \lambda x.b \ggg \lambda y.f\ x\ y = b \ggg \lambda y.a \ggg \lambda x.f\ x\ y$).

## 4 Using monads in programming practice

In computer science research, uses of monads keep relatively close to their formal definition and their original purpose. In programming practice, some uses are also straightforward application of the original definitions, albeit not for *reasoning* but for *implementing*. Other uses stretch the original concept in new interesting ways.

In this section, we look at some of the uses that are further away from the original definition. Those examples illustrate how programming concepts evolve. The monad of 2017 is not the same as the monad that was introduced into programming more than a quarter of a century ago. This is not because monads are used *incorrectly*, but because such evolution is a normal part of the life of programming concepts.

### 4.1 Monads in programming languages

Moggi [40] realized that monads provide a common structure for talking about a number of notions of computations involving, among others, state, exceptions and non-determinism. This means that some reasoning about programs can be done just using the common monadic structure, but reasoning about concrete effects still needs to involve the concrete notion of computation.

Translated to programming, monads were used to encode different notions of computations and the common structure allows code reuse. However, monads also became popular for sequencing computations and syntactic support for writing code using monads opened yet another direction.

**Code reuse** The first practical use of monads is for code reuse. In Haskell and other languages with higher-kinded types [59], it is possible to write code that works for any monad, provided that the implementation needs only *unit* and *bind*, written as return and $\gg\!\!=$, respectively. For example:

```
mapM :: Monad m ⇒ (a → m b) → [a] → m [b]
mapM f x = loop x []
  where loop [] acc = return (reverse acc)
        loop (x : xs) acc = f x ≫= λy. loop xs (y : acc)
```

The function takes a list of values $[a]$ and a function that returns $m\,b$ for each value of $a$. It applies this function to all list elements and concatenates the results, composing individual monadic values using $\gg\!\!=$.

An interesting aspect of the code is that we know very little about what the function does for a concrete monad, to the extent that the name mapM may not be appropriate in a more specific context. It is also difficult to describe the function behavior without a reference to a concrete metaphor for thinking about monads – the above description achieves this, but at the cost of clarity. More natural description might say that the function $a \to m\,b$ returns a value wrapped in a monad (using the container metaphor) or that mapM composes monadic computations sequentially (using the computation metaphor). The metaphors are almost necessary to talk about the code.

**What we talk about when we talk about monads**

**Sequencing effects**   The embedding of monads in Haskell can be used to define an order of evaluation (in an otherwise lazy language). This aspect has been utilized by the IO monad [47] to allow imperative programming and working with input/output. This also inspired the 'do' notation for working with monads in Haskell:

```
main :: IO ()
main = do
  putStr "What is your name?"
  n ← readLn
  putStr ("Hello " ++ n)
```

The 'do' notation (which has been described in academic literature only after becoming widely used) is a simple syntactic sugar that allows writing a sequence of operations and inserts monadic bind $\gg\!\!=$ at appropriate places to ensure sequencing.

The 'do' notation can be used with any monad, but it is closely linked with the metaphor that treats monads as computations. It has been criticized for making functional code look as imperative and also for making monadic style of programming preferable for syntactic reasons [5].

It is worth noting that the IO monad is, arguably, not a monad [42] in the sense that it is impossible to say whether it satisfies the monad laws. Without going into the full detail, the problem is that IO is an abstract data type (defined externally) and does not have a clear semantics for equality.

**Non-standard computations**   The IO monad pioneered the use of monads for writing non-standard computations. In case of IO, the non-standard aspect is that the computation can perform effects, which is normally not possible in Haskell.

Other languages introduced other kinds of non-standard computations. For example, F# computation expressions [46], a mechanism similar to the 'do' notation has been introduced in order to simplify writing of asynchronous computations [52]. The non-standard aspect here is that the computation can perform asynchronous I/O without blocking the current thread. For example:

```
let getLength url = async {
  try
    let! html = downloadAsync url
    return html.Length
  with e →
    return 0 }
```

The let! construct is similar to ← in the 'do' notation and return is a keyword for the *unit* operation, but the computation expression syntax takes the idea of defining non-standard computations based on monads further in two ways.

First, if we ignore the async {...} block and replace let! with let, then the code is ordinary synchronous F#. Rather than being a new notation for sequencing computations, the notation lifts ordinary F# computation into a non-standard one. Second, the try block (for exception handling) is an example of a keyword that can be exposed by providing additional operations beyond *unit* and *bind*. Monads are still an essential part of the feature, but F# stretches into yet another direction. Unlike Haskell, F# also does not have an abstraction capturing monads in general (such as the `Monad` type class), but instead requires developers to define concrete *computation builderes* such as async {...} for each concrete monad they intend to use. In a way, monads exist in the F# specification, but not as a first-class language entity.

Another instance of using monads to embed non-standard computations into a programming language is LINQ in C# and VB.Net [39]. The syntax used by LINQ is more akin to SQL. Like F#, it offers more keywords enabled by providing additional operations. Curiously enough, the *unit* operation is often ignored when talking about LINQ because it has no special syntax.

**Syntactic sugar** Both Haskell 'do' notation and F# computation expressions are syntactic sugar, introduced primarily to provide nicer syntax when using monadic abstractions. Yet, in both languages, the notation can be, and has been, used by libraries that share little with the original monadic structure. For example, the Blaze library [49] allows users to compose HTML using the 'do' notation:

```
sayHello :: String → Html
sayHello name = H.body $ do
  H.h1 "Welcome"
  H.p ("Hello " ++ name)
```

Here, the computation constructs a `Html` value and the 'do' notation is used just to concatenate multiple HTML elements. The `Html` type is not a generic type $Ma$ as required by the monad structure and it lacks the *bind* operation. A new version of the library later appeared that changes the structure to provide the monadic interface, not because the library did not *work well*, but to allow composition with other monads.

The example shows that syntactic sugar originally designed for monads can later be used in a way that has very little to do with the original idea. In the next section, we briefly reflect on the evolution illustrated by this and the previous examples.

### 4.2 How programming concepts evolve

Mathematical concepts such as *polyhedron* or *function* may appear timeless, but if we look at their history, it turns out that they evolve and definitions change to accommodate or exclude corner cases and newly discovered instances.

Lakatos [28] demonstrates the process in mathematics using polyhedra and Euler's formula ($V - E + F = 2$) that relates the number of vertices ($V$), edges ($E$) and faces ($F$) of a polyhedron. The history of the formula is a process of *proofs and refutations*. A counter example that does not satisfy the formula (such as a polyhedron with a tunnel) is used to revise the definition in order to recover the proof.

The path from category theoretical structure to a programming construct documented in Section 2 and Section 4.1 shows that programming concepts, such as monads, undergo a similar development. There is one interesting difference from mathematics though. In mathematics, the evolution is driven by aims to prove or disprove properties of an entity. In programming, there is a multitude of driving forces including formal proofs using a mathematical model, the appearance of different implementations in various programming languages and also shifts in what metaphors are used to think and talk about the concept.

More generally, the meaning of a programming language concept is constituted at three interlinked levels. For monads, the three levels provide the following:

- **Formal level.** A monad has a category theoretic definition. This is used in proofs, such as when reasoning about monadic code [15], but it also interacts with the two other levels. Examples used at the formal level contribute to the metaphorical understanding of what monads are, often because formal examples are shorter. Properties that hold at the formal level can also influence the implementation level, for example, when we show that a monad defined using *bind* is equivalent to a monad defined using *join* and *map*.

- **Implementation level.** Monads have been implemented as a type class definition (in Haskell), syntactic sugar (in F# and LINQ) and as libraries in many other languages. Unlike with Lakatos' polyhedra, the uses that do not fit the formal definition have not (yet?) led to a revision of the definition, but monad comprehensions [16] led to a new way of syntactic reasoning about queries at the formal level [18]. The 'do' notation introduced the idea of understanding monads as computations and the F# implementation of the Maybe monad led to the development of the powerful railway track metaphor (Section 2.4).

- **Metaphorical level.** The influence of the metaphorical level is more difficult to trace as it is rarely documented in writing. However, thinking of monads as computations is likely the reason why most formal definitions now prefer *bind* over *map* and *join* (the former is more convenient for programming, but the latter is simpler for proving). Similarly, thinking of monads more generally as non-standard computations likely contributed to their introduction in F# [46]. As an impure language, F# does not need monads for implementing state or exceptions, but they become useful for asynchronous computations [52].

As discussed in Section 3, monads are rooted in the mathematical Algol research programme and so it is perhaps not surprising that the core of the evolution stays around the original formal definition. However, we can see that the concept develops.

Monads turn from a formal entity inhabiting the mathematical world into the physical world of programs. This changes their purpose from reasoning to implementing. As new ways of thinking about monads appear, they become useful for solving new kinds of programming problems. This then leads to development of mathematical tools for using monads as a syntactic entity rather than category theoretical notion.

The three different levels exist for other programming concepts including types, functions, processes or objects, but they interact differently. For example, several papers analyze the concept of type [26, 38, 44]. Using our three-level framework, the analyses suggest that types appeared independently at the formal and implementation levels [26], influenced by the ordinary English-language interpretation of the word 'type' [38] at the metaphorical level. Their implementation enabled features such as auto-completion [44], which then had effects on both formal (new kinds of reasoning) and metaphorical (types as real-world entities) levels.

## 5  Monads and the programming community

In the previous section, we describe the meaning of monads at three different levels: formal, implementation and metaphorical. In this section, we briefly consider one additional aspect – the social side of monads. The main focus of this paper is on the three aforementioned levels of meaning, but we find it necessary to briefly explore the social side, as it asks important questions and offers additional references for the discussion of arguably questionable uses of monads in Section 6.

### 5.1 Monads are just monoids in the category of endofunctors

The opening quote of the paper is interesting for one more reason not mentioned before. *"A monad is just a monoid in the category of endofunctors. What is the problem?"* has become a *cultural artifact* of the programming language community. Search for the quote online and you can buy t-shirts and coffee mugs with it.

Interestingly, the quote means two exactly opposite things to two groups of people. For some, it embodies the elegance of the structure. A simple category theoretical definition leads to a programming construct that is useful in practice. For others, it symbolizes the elitism of incomprehensible "ivory tower" approach to programming.

Learning monads and writing a monad tutorial or writing a talk on monads has also become an important milestone in learning about theory of functional programming. This is illustrated by the growing number of monad tutorials recorded on the Haskell Wiki [6], although the list is now largely incomplete. Learning monads seems to have an aspect of what anthropologists call a *rite of passage*.

### 5.2 Sociology of monads

To shed some light on the social and cultural importance of monads, we consider two points made about science and mathematics. As noted in Section 3, any science is grounded in a certain research programme that provides a number of basic assumptions that proponents of the research programme do not question.

As a developer familiar with imperative programming who is learning pure functional languages, you need to change some of your basic assumptions about programming. The initial step is not, and cannot be, a rational process. We need to accept a different way of thinking and only then we can use rational reasoning to explore what follows from the assumptions. Feyerabend [11] looks at this change, analysing how old *interpretations*, which contradict new theory, are changed to allow the new theory to be accepted:

> *The offensive interpretations are replaced by others, [using tactics in which] propaganda and appeal to distant, and highly theoretical, parts of common sense are used to defuse old habits and to enthrone new ones.*

As a concrete case, consider using input and output with and without monads. Initially, using monads will certainly appear cumbersome to the uninitiated. When explaining a monadic "Hello world" program, one needs to appeal to more distant benefits of monads. Some of those benefits may be practical, such as those discussed in Section 4.1, but the mathematical roots of monads provide another appealing argument on its own. As noted by Hacking [19] who discusses experimentalism in sciences such as physics, *"we find prejudices in favor of theory, as far back as there is institutionalized science"*. The simple fact that monads are rooted in category theory may be a factor that contributes to their popularity and social attractivity.

A concrete mechanism of how this works has been described by Lakoff and Núñez [31]. Their cognitive science research looks at how abstract mathematical ideas arise from everyday embodied experience. This contrasts with what they call *The Romance of Mathematics*—a belief that mathematics is an objective feature of the universe that leads to the impression that *"mathematicians are the ultimate scientists, discovering absolute truths, not just about the physical universe, but about any possible universe"*:

> *The Romance of Mathematics makes a wonderful story. (…) It has attracted generations of young people to mathematics. [We want to believe] that, at least in doing mathematics, we can be rational, logical, and certain of our conclusions. But sadly, for the most part, it is not a true story.*

Just like attracting generations of young people to mathematics, The Romance of Mathematics attracts people to the mathematically rooted approach to programming. However, Lakoff and Núñez also describe some of the negative effects of The Romance of Mathematics. It contributes to a culture that *"rewards incomprehensibility, in which it is the norm to write only for an audience of the initiated"* [31] and it leads to *"alienation of other educated people from mathematics"* [31]. The aforementioned slogan about monads certainly fits this description. The audience of initiated sees it as elegant, while it causes an alienation of the educated, but uninitiated audience.

## 6 Undesirable uses of monads

What was said so far in this paper might be of interest as a philosophical and historical analysis of an influential concept in programming languages, but there were no takeaways for practical programming or computer science research. The aim of this section is to change that.

Despite the fact that negative results are rarely reported in academic literature, there are a number of cases where monads were used in an academic paper and their use was later revised or avoided. In this section, we review three such cases. For each of them, we suggest that treating monads in a more comprehensive way and considering the formal, implementation and metaphorical level could have prevented the undesirable use of monads.

### 6.1 Structuring the semantics of dataflow languages

The first case we consider is the use of monads for giving the semantics of dataflow programming languages such as Lucid. In dataflow programming, programs are written as computations over streams. For example $(x + \text{prev } x)/2$ creates a stream where each value is calculated as the average between the current and the previous value of the stream represented by the variable $x$.

**History of dataflow semantics**   Orchard [43] discusses the history of the semantics of dataflow programming languages which was first written using monads and later revisited using comonads, category theoretical dual of monads:

> *In 1995, Wadge proposed that the semantics of the dataflow language Lucid, which can be understood as an equational language for infinite streams, could be structured by a monad [56]. Ten years later, Uustalu and Vene gave a semantics for Lucid in terms of a comonad, and stated that "notions of dataflow cannot be structured with monads" [55]. There is an apparent conflict, which raises a number of questions.*

To summarize Orchard's account, a stream Stream $a$ can be seen as an (infinite) sequence of $a$ values. The monadic semantics of the above expression $(x + \text{prev } x)/2$ is a function of type Stream $a \to$ Stream $a$, i.e. a function that takes a stream as an input and produces stream as an output. The core of the argument against using monads in this case is that monadic computations (as we've seen in Section 2.3) are structured as functions of type $a \to M\ b$ and the monad provides the essential operation that lifts those into composable functions $M\ a \to M\ b$.

In case of monadic semantics of Lucid, we already construct composable functions of the shape $M\ a \to M\ b$ ourselves. The operations of a monad are not used to provide plumbing for composition, but instead, they are used to give semantics of some of the Lucid operations. For example, consider the *latest* operation:

> latest : Stream (Stream $a$) $\to$ Stream $a$

The operation takes a stream of streams and produces a new stream that returns the current value of the current stream at each point in time. The first value will be the first value of the first stream, the second value will be the second value of the second stream, etc. This is a useful Lucid combinator and it has the same type as monadic *join*. However, unlike other uses of monads we have seen, such *join* operation does not provide a way of composing computations (in category theoretical terms, the computation is not captured by a Kleisli morphism). This has a number of consequences, as noted by Orchard [43]:

> *A computation solely captured by Kleisli (...) morphisms benefits from equational laws for reasoning and optimization by simplification, as well as better syntactic support in languages. Such benefits are not available for [morphisms that do not follow this structure].*

The solution proposed by Uustalu and Vene [55] is to structure the semantics using a comonad that provides a way of composing computations of type $C\ a \to b$. This fits with dataflow programming, because we can structure many operations as functions Stream $a \to b$ that takes a stream (with current value and history) and computes the new current value.

**Rethinking the use of monads**   Does the discussion in this paper provide any hint that the use of monads for the semantics of dataflow is not desirable? The use of monads seems appropriate at the implementation level – we can define operations of the appropriate type and use them for writing dataflow programs. However, potential issues become apparent when we consider the formal and metaphorical level.

The formal aspects have been analyzed in detail by Orchard [43]. Although we can define a monad, the way we use it is not suitable for formal reasoning using Kleisli arrows. At the metaphorical level, a function of type $a \to \mathsf{Stream}\ b$ can be seen as a producer of a stream, but not as a dataflow computation itself. We can see $\mathsf{Stream}\ a$ as a container (containing present and past values), but the latest operation cannot be easily interpreted as "unwrapping" a nested container; latest must throw away some values, which is not the case when unwrapping a nested container.

### 6.2 Parsers and a tempting abstraction

The second case we consider is the use of monads in parsing. The early work on parser combinators [23] notes that there is a relationship with monads. Parsers were also mentioned by Wadler [57] and more work on monadic parsing soon followed [24].

**Understanding monadic parsers**  A parser $\mathsf{Parser}\ a$ is a function $\mathsf{String} \to (a, \mathsf{String})$ that takes an input string and returns a parsed value of type $a$ together with the remaining unconsumed input. Parser combinators provide ways of combining parsers. We can define a *choice* combinator of type $\mathsf{Parser}\ a \to \mathsf{Parser}\ a \to \mathsf{Parser}\ a$ that takes two parsers and returns a new parser that succeeds if either of the two parsers succeed. Interestingly enough, it is also possible to provide the following two combinators:

$$
\begin{aligned}
bind\ &::\ \mathsf{Parser}\ a \to (a \to \mathsf{Parser}\ b) \to \mathsf{Parser}\ b \\
unit\ &::\ a \to \mathsf{Parser}\ a
\end{aligned}
$$

The *unit* operation creates a parser that always succeeds without consuming any input. The *bind* operation creates a parser that applies the first parser and, if it succeeds and produces a value of type $a$, calls the function to construct a parser to parse the rest of the input. This implements the monadic interface and so it becomes possible to use parsers with the 'do' notation. Swierstra and Duponcheel [51] later noted that the monadic parsers have several drawbacks and developed a way of fixing those:

> *The normal disadvantages of conventional [monadic] parsers, such as their lack of speed and their poor error reporting are remedied.*

Their new approach provides an efficient implementation with better error reporting, but it cannot define the *bind* operation and thus is not monadic:

> *The techniques [do not] extend to monad-based parsers. [T]he monadic formulation [causes] the evaluation of the parser construction over and over (...).*

One of the problems is that the parser needs to repeatedly call the function $a \to \mathsf{Parser}\ b$ to construct a parser. This also means that the parser for the second part of the input is only known after parsing the first part of the input. For context-free grammars, this is not necessary as the structure of the parser does not depend on the parsed values. Swierstra and Duponcheel replace the *bind* operation with an operation of type $\mathsf{Parser}\ a \to \mathsf{Parser}\ b \to \mathsf{Parser}(a \times b)$ which sequentially composes two parsers. Here, the second parser does not depend on the result of the first one and the combinator can be implemented more efficiently. It is tempting to implement the monadic interface for parsers, because it is well-defined and relatively easy to provide, but this makes it impossible to discover a more efficient implementation of parser combinators.

**Rethinking the use of monads**   How could the understanding of cognitive and social aspects of monads prevent the undesirable use of monads in this case? There are two possible answers. First, the social side of monads explains why monads are a tool that one might want to use without considering whether it is needed. The *bind* combinator in the early papers was not necessarily useful, but it was implemented nevertheless, to show that parsers form a monad. Thus, being aware of the Algol research programme might make us reconsider whether *bind* provides merely an interesting result at the formal level, or whether it is genuinely useful at the implementation level.

Second, we could imagine an understanding of monads that more strongly emphasizes the metaphorical level. Many of the standard, widely used monads fit one of the two metaphors we introduced in Section 2.4, but it is not obvious how to interpret parsers as either containers or as non-standard computations (or railway tracks). As noted earlier, the metaphors are not just useful for teaching. They capture the well-understood and tested use cases of monads.

### 6.3  Monad as the uninteresting part

The last case of controversy about monads in computer science literature that we consider in this paper is the use of monads for concurrent and parallel programming. An example of work in this area is the Par monad [37]. The paper introduces a type Par $a$ that represents a computation which may involve running work in parallel.

**Understanding parallel computations**   The Par monad represents a computation, so it fits with one of the monad metaphors. It also implements both of the operations required by a monad and the implementation satisfied the required laws:

$$\begin{array}{lll} bind & :: & \text{Par } a \to (a \to \text{Par } b) \to \text{Par } b \\ unit & :: & a \to \text{Par } a \end{array}$$

The *unit* operation creates a computation that immediately returns the given value. The *bind* operation sequentially composes the first computation with the one generated by the given function.

**Rethinking the use of monads**   An objection against using monad in this case is not that it is inappropriate, but that it is the least interesting thing about the abstraction. Monad provides a way of sequentially composing computations, but what really matters about Par $a$ is what kind of parallelism and synchronization primitives it provides (forking and write-once shared variables). The title of the paper, "A monad for deterministic parallelism" [37] might suggest that a monad is used for deterministic parallelism, but the parts of the abstraction that provide parallelism are everything *except* for the monadic part, which is there to allow sequential composition.

At the metaphorical level, we can see that the Par monad is a computation and we can also understand that sequencing is the least interesting part. The main purpose of the Par monad is to synchronize the complex system of railway tracks, rather than link them. This idea has been captured, for example, by the work on joinads [45], which extends monads (and the `MonadPlus` type class [54]) with the following operations:

$$\begin{array}{lcl} \mathit{merge} & :: & m\ a \to m\ b \to m\ (a,b) \\ \mathit{choose} & :: & m\ a \to m\ a \to m\ a \end{array}$$

The two operations provide additional way in which computations, or *railway tracks*, can be composed; *merge* joins two computations and returns a pair with both individual results and *choose* implements a non-deterministic choice.

Why is the rhetorical focus of the paper about the Par type on the monadic abstraction? This is, most likely, the result of the social side of monads. They are a topic of interest for the community (Section 5) and they became a key tool of the purely functional research programme (Section 3.2).

## 7 A case for wider understanding

This paper looks at the programming concept of a monad from a wider, historical, philosophical and cognitive perspective. The method we use is to consider the meaning of a programming concept at three different levels: the *formal level* used for proving mathematical properties about the concept, the *implementation level* at which concepts are used for writing concrete code and the *metaphorical level* which is used for intuitive thinking and explaining the concept.

### 7.1 Future and related work

The idea of treating programming concepts as entities consisting of multiple levels is not new. Turner and Angius [53] treat programs as technical artifacts with two levels, one is a specification and function and the other is an implementation. Our analysis differs in that we look at *programming concepts* rather than programs as a whole and we separate the formal and intuitive aspects of the specification.

We briefly hinted how similar analysis might apply to other programming concepts such as types in Section 4.2, but we believe that doing an in-depth analysis of a number of other concepts can help us better understand how programming concepts evolve *in general*. For example, looking at types suggests that the interaction pattern between the three levels that we uncovered for monads is just one particular kind.

The other interesting problem is finding good ways of analysing individual levels. Theoretical computer science provides a rich set of resources for looking at the formal level. We also have ways of talking about the implementation level, but often do so indirectly via the formal level. For the metaphorical level, this paper referred to relevant work in cognitive sciences, but we believe there is much left to be done in understanding how we use metaphors when programming.

There are also numerous uses of monads that we were not able to cover in this paper such as the link between monads and aspect-oriented programming [8, 32]. This relationship might provide yet another metaphor for thinking about monads that is perhaps equally interesting as the two metaphors, containers and computations, presented in this paper.

### 7.2 Summary – What is a monad?

The short answer to the question *"What is a monad?"* is that it is a monoid in the category of endofunctors or that it is a generic data type equipped with two operations that satisfy certain laws. This is correct, but it does not reveal an important bigger picture. This is because the question is wrong. In this paper, we aim to answer the right question, which is *"What do authors really say when they talk about monads?"*

Our answer situates monads within the disciplines of cognitive science, philosophy of science and mathematics. We looked at the metaphors that are used when teaching monads and discussed why these should be seen as an inherent part of the answer. We looked at how monads are used in academic research and in practice. This illustrates how programming concepts evolve and reveals interactions between the formal, implementation and metaphorical levels. In the case of monads, the formal definition remains stable, but our intuitive understanding of the concept and how we use it in programming changes over time.

Finally, we discussed why such broader understanding of programming concepts such as monads is useful. We looked at a number of cases, documented in the literature, where a monad was used in a way that was revised in later work. For each case, we suggested that a broader understanding of monads could help the original authors to avoid the issues discovered later.

**Acknowledgements**   The anonymous reviewers provided fantastic feedback that helped me to better frame the paper and significantly improve many of the important sections. I would like to thank to fellow members of the Revolution & Beers Cambridge group, namely Dominic Orchard, Sam Aaron, Antranig Basman and Stephen Kell. I would also like to thank to Giacomo Citi for a QuickCheck reference and Torsten Grust for numerous corrections. This work was in part supported by The Alan Turing Institute under the EPSRC grant EP/N510129/1.

**What we talk about when we talk about monads**

## About the author

**Tomas Petricek** Tomas is a Visiting Researcher at the Alan Turing institute, working on tools for open data-driven storytelling (http://thegamma.net). His many other interests include programming language theory (his PhD thesis is on *coeffects*, a theory of context-aware programming languages), open-source and functional programming (he is an active contributor to the F# ecosystem), but also understanding programming through the perspective of philosophy of science. Contact him at tomas@tomasp.net.

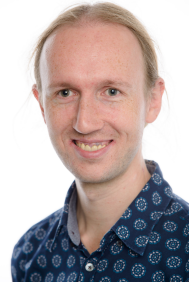